\documentclass[baaa]{baaa}

\usepackage[pdftex]{hyperref}
\usepackage{subfigure}
\usepackage{natbib}
\usepackage{helvet,soul}
\usepackage[font=small]{caption}
\usepackage{multirow}
\usepackage{soul}
\usepackage{xcolor}
\sethlcolor{lime}

\usepackage{upgreek}

\contriblanguage{1}

\contribtype{1}

\thematicarea{5}

\received{\ldots}
\accepted{\ldots}

\title{Exploring the chemical evolution in hot molecular cores}

\titlerunning{Chemistry of hot molecular cores}

\author{N.C. Martinez\inst{1,2}, S. Paron\inst{1}, M.E. Ortega\inst{1}, L. Sup\'an\inst{1} \& A. Petriella\inst{1}
}

\authorrunning{Martinez N. C. et al.}

\contact{nmartinez@iafe.uba.ar}

\institute{
Instituto de Astronom{\'\i}a y F{\'\i}sica del Espacio, CONICET--UBA, Argentina
\and
Universidad de Buenos Aires. Facultad de Ciencias Exactas y Naturales. Departamento de F{\'\i}sica. Buenos Aires, Argentina
}

\resumen{Presentamos los resultados preliminares de un extenso proyecto de investigación que apunta a describir las condiciones físicas y químicas de núcleos moleculares calientes (HMCs, por sus siglas en inglés). Utilizando datos espectroscópicos y del continuo milimétrico extraídos del archivo del Atacama Large Millimeter Array (ALMA) se han estimado temperaturas de rotación ($\rm T_{rot}$) y densidades de columna de $\rm{CH_{3}CN}$, $\rm{CH_{3}CCH}$, y A-- y E--$\rm CH_{3}OH$ para una muestra de núcleos moleculares. Se presenta una caracterización térmica de dichos núcleos, revelando la existencia de gradientes de temperatura dentro de los mismos. Estos núcleos se encuentran embebidos en nubes moleculares extensas. Adicionalmente se estimaron abundancias moleculares que fueron estudiadas como trazadoras de la evolución química de dichos núcleos. Finalmente, como estudio piloto para interrelacionar observaciones y simulaciones, se comparan algunas de las abundancias moleculares obtenidas con simulaciones generadas con el código Nautilus.}

\abstract{We present preliminary results of an extensive research project aimed at describing the physical and chemical conditions of hot molecular cores (HMCs). Using millimeter continuum and spectroscopic data extracted from the Atacama Large Millimeter Array (ALMA) archive, we have estimated rotational temperatures ($\rm T_{rot}$) and column densities of $\rm{CH_{3}CN}$, $\rm{CH_{3}CCH}$, and A-- and E--$\rm CH_{3}OH$ for a sample of molecular cores. We present a thermal characterization of these cores, revealing the existence of temperature gradients within them. These cores are, in turn, embedded in large molecular clouds. Additionally, we estimated molecular abundances that were evaluated as tracers of the chemical evolution of these cores. Finally, in a pilot study aimed to link observations with simulations, some of the obtained molecular abundances are compared with predictions from the Nautilus code.}


\begin{document}

\maketitle
\section{Introduction}
\label{intro}

Hot molecular cores (HMCs) are dense gaseous structures with sizes of about 0.1 pc. They are the chemically richest structures in the interstellar medium (ISM) and, therefore, the primary astrochemical laboratories for investigating how complex molecules form in space (e.g., \citealt{herbst09,jor20,coletta20}). Given that such cores are the sites where massive stars are born, they represent the most appropriate places in which to investigate the chemistry related to massive star formation. Indeed, the molecular matter is present at all spatial and temporal scales related to star formation \citep{jor20}.

Methyl cyanide (CH$_3$CN) and methyl acetylene (CH$_3$CCH) are usually used to estimate the gas temperature through the rotational diagram (RD) method \citep{goldsmith99}. From such a method, the rotational temperature ($\rm T_{rot}$) and molecular column densities of these species can be obtained to investigate the thermal conditions of the places where such molecules are detected. Certainly,  CH$_3$CN and CH$_3$CCH have been extensively studied towards many hot molecular cores \citep{remijan04,calcutt19,brouillet22,ortega23}.
Actually, any molecule that possesses several rotational transitions, or K-ladder transitions, with well-separated energy levels and whose emissions fall into narrow and easily observable spectral ranges, can be used in RDs to estimate temperatures.

Considering that the emission of each molecule can originate from different gas layers within a molecular core, performing the mentioned process with multiple species towards a sample of molecular cores is useful for statistically studying temperature gradients in such dense gaseous structures related to star formation. In this brief article, we report preliminary results of an ongoing research project, conducted over a sample of 37 molecular cores using several molecular species. Here, we present some results obtained from methyl cyanide (CH$_{3}$CN), methyl acetylene (CH$_{3}$CCH), and methanol (CH$_{3}$OH) in some cores of the mentioned sample.

\section{Data and sample of sources}

In this study, we analyse 10 HMCs from the Atacama Large Millimeter Array (ALMA) survey project 2015.1.01312.S (PI: Fuller, G.). The data, in Band 6 (spectral range: 224.2--242.7 GHz), were obtained from the ALMA Science Archive.\footnote{http://almascience.eso.org/aq/} 
The single pointing observations for the target were carried out using the telescope configuration L5BL/L80BL(m) 42.6/221.3 in the 12~m array. The beam size of the observations is about 0$.\!\!^{\prime\prime}$7 and the line sensitivity (10 km s$^{-1}$) is 1.5 mJy beam$^{-1}$. The spectral resolution is 1.1 MHz ($\Delta$v = 1.4 km s$^{-1}$), and the field of view is 25$^{\prime\prime}$ with a maximum recoverable scale of 6$.\!\!^{\prime\prime}$1. 

The fields observed in project 2015.1.01312.S were known to host young high-mass embedded protostellar sources,
selected from both the Red MSX Source (RMS) survey and the Spitzer Dark Cloud (SDC) sample (see \citealt{avison23}).  
In this article, we present the results obtained from ten sources of the sample (see Table\,\ref{sources}). 

An inspection of the data reveals significant molecular richness in all sources. From this diverse inventory, we selected four species with sufficient transitions to ensure reliable RDs. The molecules chosen are the symmetric tops CH$_{3}$CN and CH$_{3}$CCH, as well as the A/E-symmetry states of CH$_{3}$OH. Transitions, rest frequencies ($\nu_{\rm rest}$), and energy of the upper levels (E$_{\rm u}$) of the used molecular lines are presented in Table\,\ref{transitions}.

\begin{table}
\caption{Sample of studied sources.}
\centering
\tiny
\begin{tabular}{lllr}
\hline\noalign{\smallskip}
\multirow{2}{*}{\#} & \multirow{2}{*}{ALMA source name} & \multicolumn{2}{c}{Coordinates} \\
\cline{3-4} \noalign{\smallskip}
& & R.A. (J2000) & Dec. (J2000) \\
\hline\noalign{\smallskip}
1 & G029.8620$-$00.0444	&	18:45:59.6	&	-02:45:07.0	\\
2 & G326.6618$+$00.5207	&	15:45:02.8	&	-54:09:03.0	\\
3 & G332.9868$-$00.4871	&	16:20:37.8	&	-50:43:50.0	\\
4 & G339.6221$-$00.1209	&	16:46:06.0	&	-45:36:44.0	\\
5 & SDC20.775$-$0.076	&	18:29:12.2	&	-10:50:35.0	\\
6 & SDC24.462$+$0.219	&	18:35:11.6	&	-07.26:23.0	\\
7 & SDC28.147$-$0.006	&	18:42:42.5	&	-04:15:34.0	\\
8 & SDC42.401$-$0.309	&	19:09:49.9	&	08:19:47.0	\\
9 & SDC43.186$-$0.549	&	19:12:09.2	&	08:52:15.0	\\
10 & SDC43.877$-$0.755	&	19:14:26.2	&	09:22:35.0	\\
\hline
\end{tabular}
\label{sources}
\end{table}

\begin{table}[h!]
\centering
\tiny
\caption{Analysed molecular lines}
\label{transitions}
\begin{tabular}{lcr}
\hline\hline
 & & \\[-2ex]
Transition & $\rm \nu_{rest}$ (GHz)   & E$\rm_{u}$ (K) \\
\hline
 & & \\[-2ex]
\multicolumn{3}{c}{CH$_{3}$CN v=0} \\
\hline
  13(6)-12(6) & 238.972 & 337.37\\
 13(5)-12(5) & 239.022 & 258.87 \\
 13(4)-12(4) & 239.064 & 194.62 \\
    13(3)-12(3) & 239.096 & 144.63 \\
    13(2)-12(2) & 239.119 & 108.92 \\
    13(1)-12(1) & 239.133 & 87.49 \\
    13(0)-12(0) & 239.137 & 80.34 \\
 \hline
  & & \\[-2ex]
\multicolumn{3}{c}{CH$_{3}$CCH} \\
 \hline
  14(3)-13(3) & 239.211 & 150.36 \\
  14(2)-13(2) & 239.234 & 114.67 \\
  14(1)-13(1) & 239.241 & 93.26  \\
14(0)-13(0) & 239.252 & 86.12  \\
\hline
 & & \\[-2ex]
 \multicolumn{3}{c}{CH$_{3}$OH v=0} \\
\hline
 20(2,18)-19(3,16)E & 224.699 & 514.26 \\
   21(1,20)-21(0,21)E & 227.094 & 557.07 \\
   12(1,2)-11(-2,10)E & 227.229 & 186.43 \\
   16(1,16)-15(2,13)A & 227.814 & 327.24 \\
   16(3,13)-17(0,17)E & 239.397 & 378.28 \\
   16(7,9)-17(6,11)A & 239.731 & 560.07 \\
   5(1,5)-4(1,4)A & 239.746 & 49.06 \\
   5(3,2)-6(2,4)E & 240.241 & 82.53 \\
   22(6,16)-23(5,19)E & 241.043 & 775.57 \\
   25(3,22)-25(2,23)A & 241.590 & 803.70 \\
   5(0,5)-4(0,4)E & 241.700 & 47.93 \\
   5(-1,5)-4(-1,4)E & 241.767 & 40.39 \\
   5(0,5)-4(0,4)A & 241.791 & 34.82 \\
   5(4,1)-4(4,0)A & 241.807 & 115.16 \\
   5(-4,2)-4(-4,1)E & 241.813 & 122.72 \\
   5(-4,1)-4(-4,0)E & 241.829 & 130.82 \\
   5(3,3)-4(2,3)A & 241.832 & 84.62 \\
   5(2,4)-4(2,3)A & 241.842 & 72.53 \\
   5(3,2)-4(3,1)E & 241.843 & 82.53 \\
   5(-3,3)-4(-3,2)E & 241.852 & 97.53 \\
   5(1,4)-4(1,3)E & 241.879 & 55.87 \\
   5(2,3)-4(2,2)A & 241.887 & 72.53 \\
   5(2,3)-4(2,2)E & 241.904 & 57.07 \\
   14(-1,14)-13(-2,12)E & 242.446 & 128.93 \\
   24(3,21)-24(2,22)A & 242.491 & 745.73 \\
\hline
\end{tabular}
\end{table}

\section{Results}

\begin{figure}[h!]
    \centering
     \includegraphics[width=7.5cm]{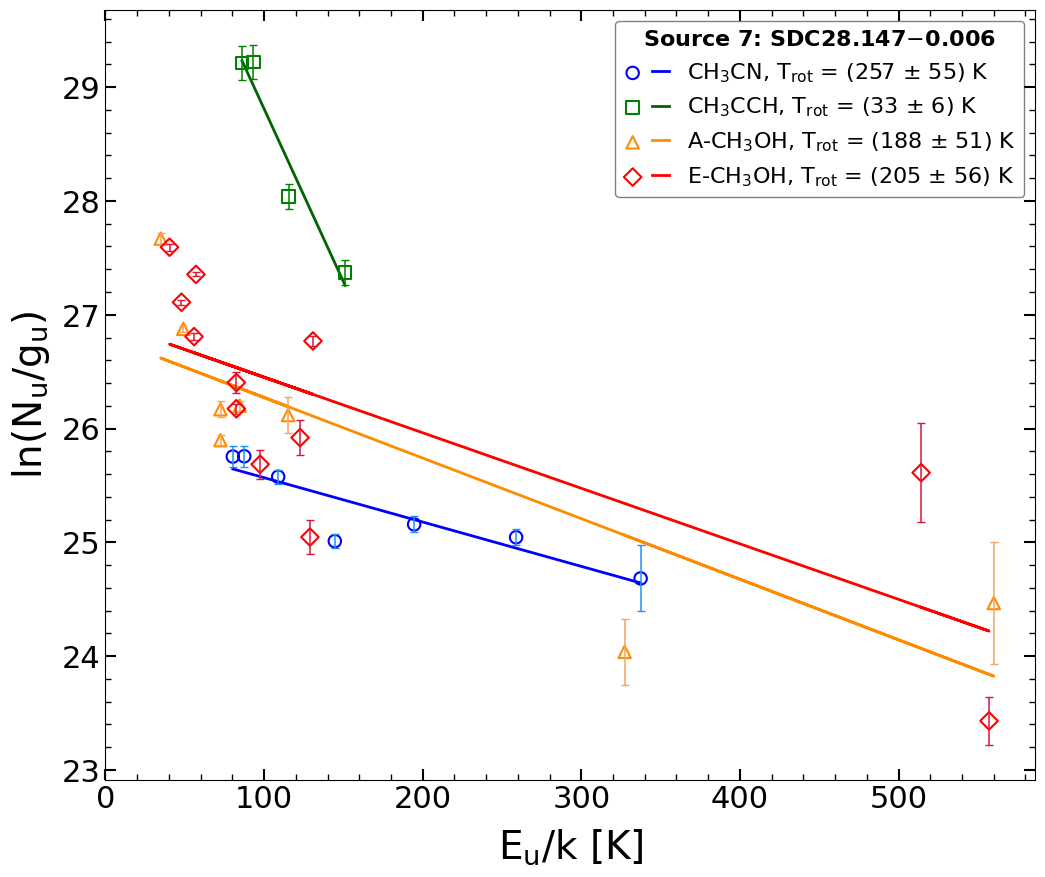}
    \caption{Combined rotational diagram used to derive the T$\rm_{rot}$ towards SDC28.147$-$0.006 (source\,7) from the four studied molecular species. The lines represent the best linear fits to the data.}
    \label{DRs}
\end{figure}

The first step in our analysis was to obtain the integrated intensities (W) for each molecular transition by fitting Gaussian profiles to the observed emission lines. With these values (not presented here due to space limitations), we then derived the rotational temperature (T$\rm_{rot}$) and the total column density (N$\rm_{tot}$) for each molecule at each core, using the RD method \citep{goldsmith99}. This technique assumes local thermodynamic equilibrium (LTE) and that the emission is optically thin. Under these conditions, one can assume that $\rm T_{rot}$ is equal to the kinetic temperature of the gas ($\rm T_{k}$) while a linear relationship exists between the logarithm of the upper-level column density (N$\rm_{u}$) and the upper-level energy (E$\rm_{u}$) as shown in the following equation:

\begin{equation}
{\rm ln\left(\frac{N_u}{g_u}\right)={\rm ln}\left(\frac{N}{Q_{rot}}\right)-\frac{E_u}{kT_{rot}}},   \label{RotDia} 
\end{equation}

\noindent where N is the total column density, g$_{\rm u}$ is the degeneracy of the upper-level energy, Q$_{\rm rot}$ the partition function, and k the Boltzmann constant. 

The first term of the previous equation can be written as a function of W and spectroscopic constants related to the molecular species (see \citealt{ortega23} for details). All constants were retrieved from Splatalogue.\footnote{https://splatalogue.online/\#/advanced}
T$\rm_{rot}$ is derived from the inverse of the slope of a linear fit to the data, and N from its y-intercept. To obtain N, the values of Q$_{\rm rot}$ used were obtained from extrapolations according to the T$_{\rm rot}$ calculated in each case. Figure\,\ref{DRs} presents a representative example of the RD analysis, showing the data and linear fits for the four analysed molecular species in the source\,7.

\begin{figure}[h]
    \includegraphics[width=8.3cm]{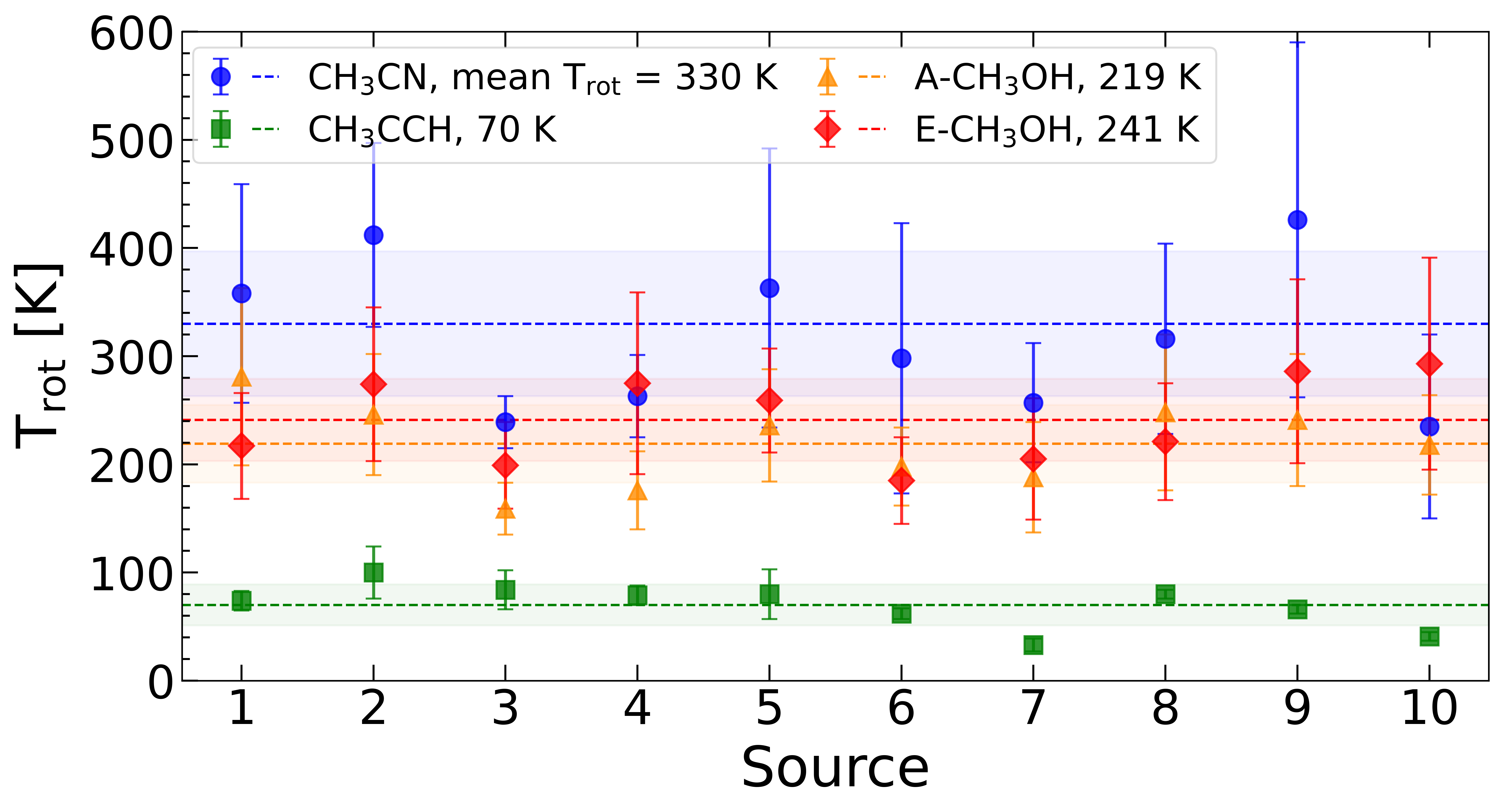}
    \caption{T$\rm_{rot}$ derived from the studied molecular species across the sample of ten regions. Horizontal dashed lines and shaded areas indicate the mean temperature and associated standard deviation ($\pm1\upsigma$), respectively, for each molecule.}
    \label{trot}
\end{figure}

Figure\,\ref{trot} presents the T$\rm_{rot}$ derived for the ten studied sources. The results reveal a clear and uniform trend throughout the sample, indicating an apparent temperature stratification within the cores. CH$\rm_{3}$CN consistently traced the hottest gas component, with a mean temperature of 330 K. In contrast, CH$\rm_{3}$CCH traced the coolest gas, with a mean temperature of 70 K. The methanol species probed warm temperatures, with mean values of 219 and 241 K for A- and E-CH$\rm_{3}$OH, respectively.

\begin{figure}[h]
\centering
    \includegraphics[width=8.2cm]{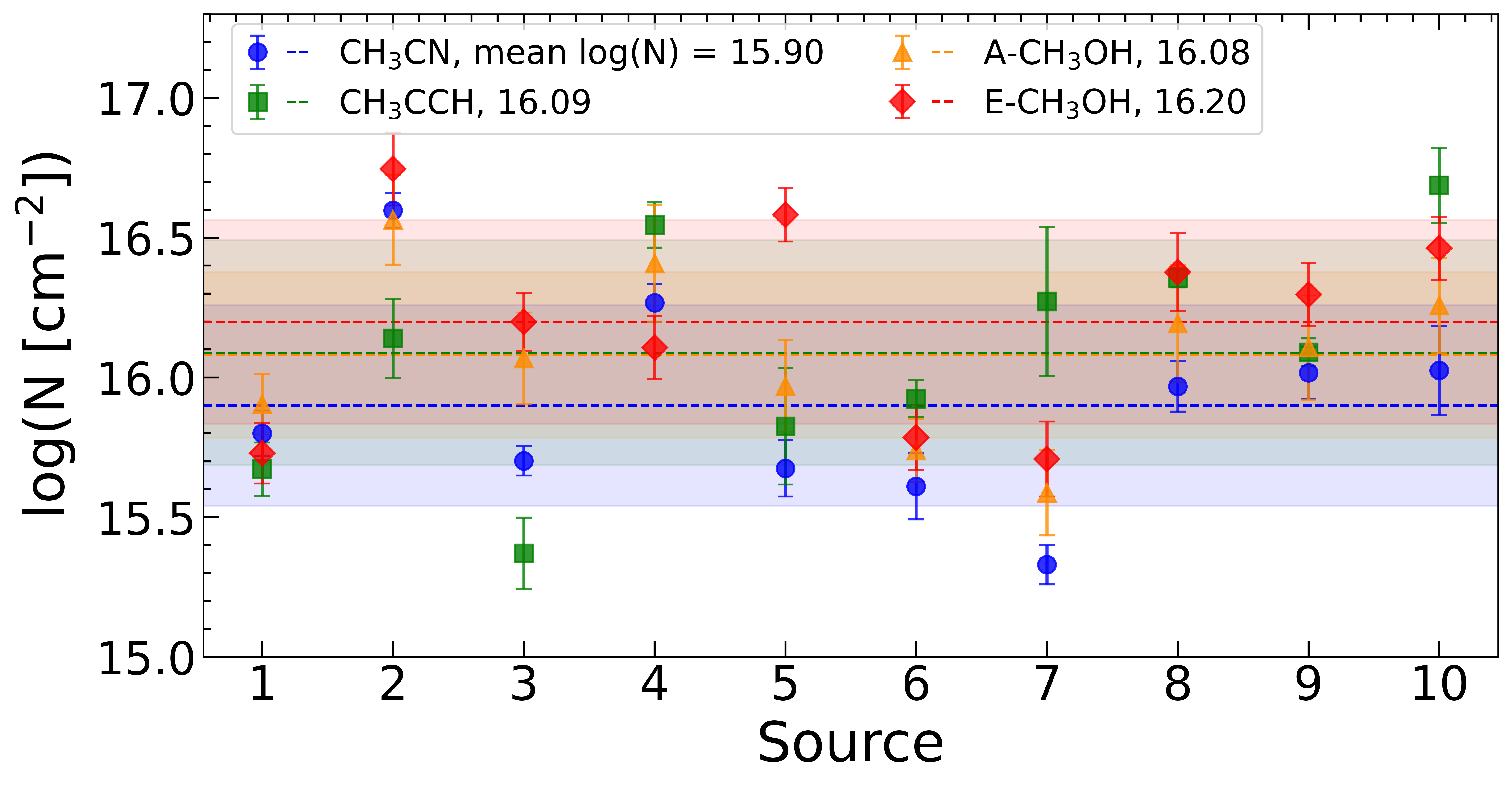}
    \includegraphics[width=8.3cm]{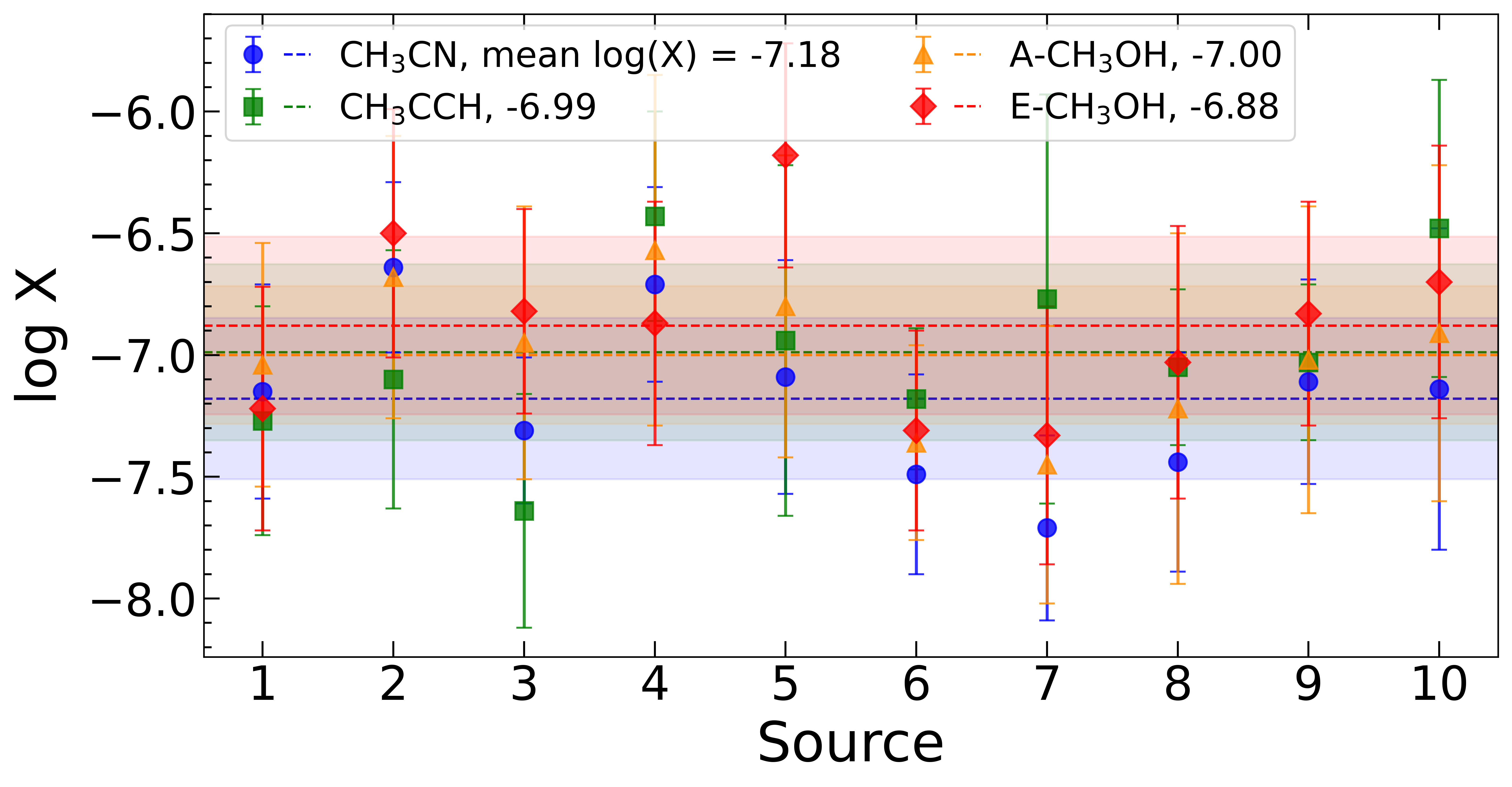}
    \caption{Obtained column densities and abundances (upper and bottom panels, respectively). Mean values are indicated with dashed lines, and shaded regions show the standard deviation around them ($\pm1\upsigma$).} 
    \label{abundances}
\end{figure}

Total column densities for each species were obtained, and subsequently, fractional abundances ($\rm X = N/N_{H_{2}}$) were calculated using the H$_{2}$ column density ($\rm N_{H_{2}}$). The $\rm N_{H_{2}}$ values were obtained for each core from the continuum emission maps, following \cite{kauffmann08}. For these calculations, it was assumed that the dust temperature is the same as the gas temperature, where the mean T$_{\rm rot}$ of the four molecular species was used in each case. The used dust opacity is 0.01 cm$^{2}$ g$^{-1}$. 
Figure\,\ref{abundances} compiles the logarithmic values of N$\rm_{tot}$ (upper panel) and the logarithmic values of X (bottom panel). 

The mean logarithmic column densities are $\rm 15.90$ for CH$\rm_{3}$CN, $\rm 16.09$ for CH$\rm_{3}$CCH, $16.08$ for A-CH$\rm_{3}$OH, and $16.20$ for E-CH$\rm_{3}$OH. The corresponding mean logarithmic abundances are  $-7.18$, $-6.99$, $-7.00$, and $-6.88$, respectively.

\begin{figure}[h!]
    \hspace{-2ex}
    \centering
    \includegraphics[width=8.3cm]{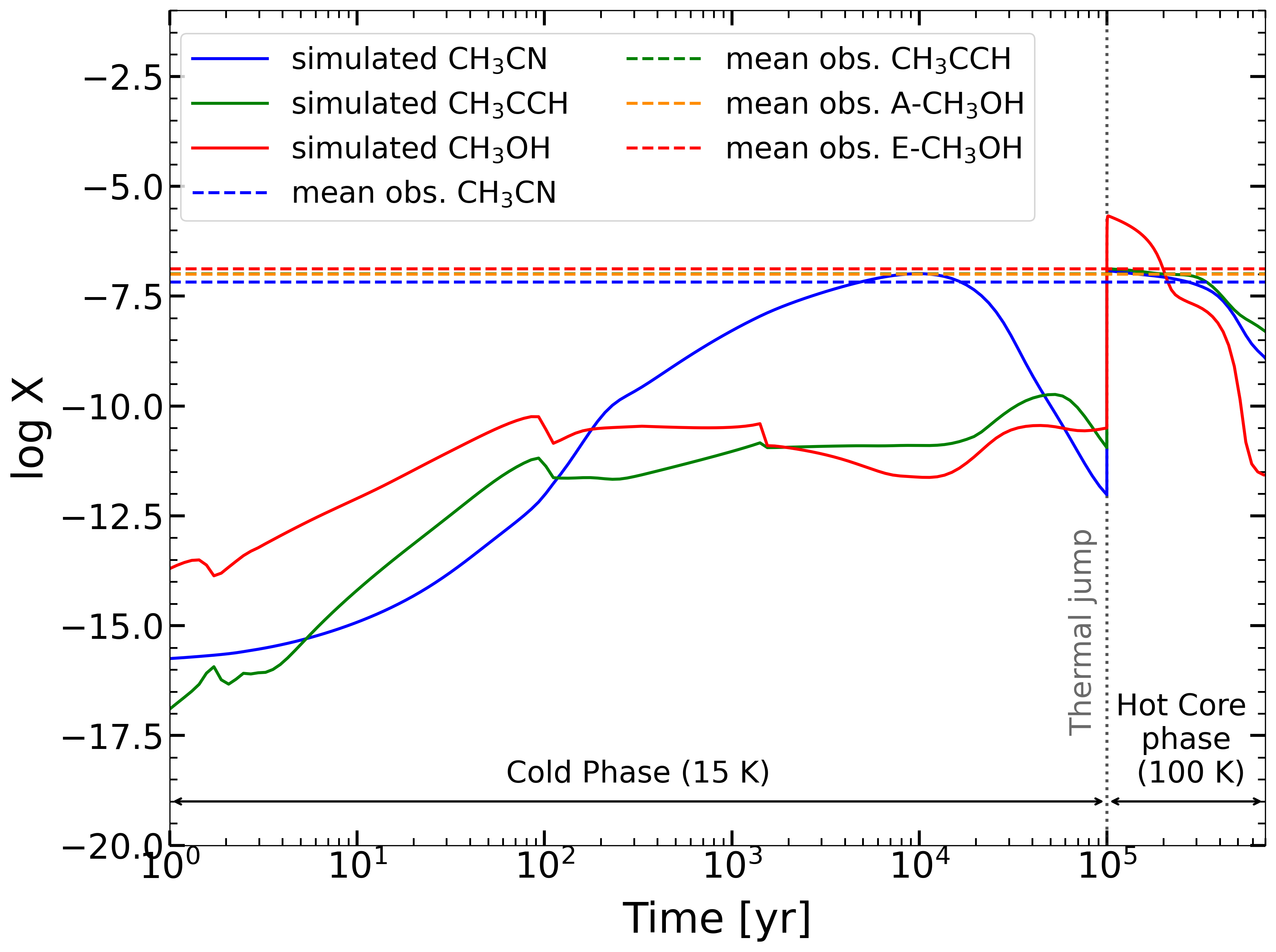}
    \caption{Chemical evolution of the studied molecular species 
    simulated with the Nautilus gas-grain code in two stages: a cold phase ($\rm 15~K$) and a hot core phase ($\rm 100~K$). The solid lines represent the model predictions for the fractional abundances. The horizontal dashed lines show the mean logarithmic abundances derived from our ten-source sample (see Fig.\,\ref{abundances}). The vertical dotted grey line indicates the thermal jump.}
    \label{nautilus}
\end{figure}

Additionally, to probe the timescale of the sources, the mean abundances were compared with the predictions of a chemical model. For this purpose, we used the Nautilus gas-grain chemical code to simulate the time-dependent evolution of chemical species in interstellar environments \citep{ruaud16}. The code
solves the kinetic equations for the public network of chemical reactions \texttt{kida.uva.2014} \citep{kida14}, considering a three-phase model that includes gas-phase chemistry and chemical processes on the surface and in the mantle of dust grains. In this work, we performed a two-stage simulation to model the chemical enrichment. The first stage simulates a cold, dense phase ($\rm T = 15~K$, $\rm n_{H_{2}} = 4\times10^{5} cm^{-3}$) from $\rm t = 0~years$ to $\rm t = 10^{5}$ years, promoting the buildup of ice mantles. We adopted an elemental ratio of $\rm C/O = 1.0$ to ensure an excess of free carbon, which is essential for the synthesis of complex organic molecules through successive surface hydrogenations. The final abundances from this phase served as initial conditions for a second, stationary hot phase ($\rm T=100~K$). This second stage evolves the chemistry from $\rm 10^{5}$ years up to a total age of $\rm 7\times10^{5}$ years, during which the thermal jump simulates the rapid sublimation of ices and subsequent gas-phase chemistry. 

The results of the chemical simulation are presented in Fig.\,\ref{nautilus}, which displays the predicted evolution of the abundances. The adopted chemical network treats methanol as a single species, without separating its A and E forms, because most kinetic databases lack spin-symmetry-specific rate reactions. Overplotted on the model tracks are the mean abundances derived from our ten-source sample. By comparing the observational data with the model, we found that the observed abundances are best reproduced at a chemical age between $2\times10^{5}$ and $3\times10^{5}$ years.

\section{Discussion}

The distinct rotational temperatures derived from the different molecular species (see Fig.\,\ref{trot}) strongly support the presence of thermal gradients within the studied HMCs. The high temperatures consistently probed by CH$_{3}$CN (T$\rm_{rot} =$ 330 K) confirm its role as a tracer of the innermost, hottest, and likely densest regions near the embedded protostar, compatible with observations of other HMCs where this molecule traces temperatures exceeding 100--200 K \citep{beltran05,araya05}. 

Conversely, the significantly lower temperatures derived from CH$_{3}$CCH (T$\rm_{rot}=$ 70 K) suggest that it primarily traces the cooler, less dense, outer envelopes. This is compatible with chemical models in which this species is efficiently formed in gas and solid phases from simpler hydrocarbon precursors such as $\rm C_2H_4$ and $\rm C_3H_5$ \citep{andron18}. 

The observed temperature stratification reinforces the physical and chemical layering expected in these evolving star-forming regions, agreeing with a structure heated from within by the central protostar \citep{kaufman97}.

Methanol (A and E isomers) tracing intermediate temperatures (T$\rm_{rot}=219 $-$ 241$ K) provides crucial insights into the interplay between gas-phase and grain-surface chemistry. Its abundance dramatically increases in the gas phase when dust grains reach temperatures sufficient for ice mantle sublimation ($\approx$ 100 K), releasing molecules formed on the surfaces \citep{bisschop07}. The temperatures we derived for methanol place it in a distinct warm zone, hotter than the CH$_{3}$CCH envelope but potentially surrounding the central  CH$_{3}$CN peak. This spatial differentiation underscores the importance of combining tracers sensitive to different temperature regimes for mapping the complex transitions between the cold envelope, the warm sublimated region, and the hot central area in HMCs.

Beyond the thermal structure, the derived molecular abundances (Fig.\,\ref{abundances}) serve as powerful diagnostics of the chemical state and evolutionary stage. These values, with mean $\rm log(X)$ ranging from $-7.18$ for  CH$_{3}$CN to $-6.88$ for E-CH$_{3}$OH, fall within the range observed in other HMCs, suggesting thar our sample is indeed representative of this class of objects \citep{gerner14}. 

To quantitatively assess the evolutionary timescale, we compared the obtained mean abundances with the model predictions (Fig.\,\ref{nautilus}). The simulation indicates that the observed abundance levels are matched by the model within a relatively narrow timeframe around $3\times10^{5}$ years, after the thermal jump that precisely takes place in HMCs. This chemical age is consistent with the dynamically inferred lifetimes of HMCs, typically estimated to be a few $ 10^{5}$ years before feedback from the central massive star(s) disrupts the core \citep{beuther25}.

In conclusion, this preliminary analysis of ten HMCs demonstrates the power of using multiple molecular tracers to reveal thermal stratification. The derived temperatures and abundances are consistent with the well-established picture of HMCs as chemically rich, centrally heated objects. Furthermore, comparison with the Nautilus chemical model provides a quantitative estimate of their chemical age, aligning with typical HMC lifetimes. While the assumptions of LTE and optical thinness inherent in the RD method warrant caution, the consistency across tracers and with models strengthens our findings. Extending this detailed analysis to the full sample of 39 sources will provide a statistically robust view of the prevalence and nature of thermal and chemical gradients during the early stages of massive star formation.

\begin{acknowledgement}
We thank the anonymous referee for her/his valuable comments.
N.C.M. is extremely grateful to the LOC for the support received to attend the 67$\rm^{a}$ Reunión de la Asociación Argentina de Astronomía at Mendoza, Argentina. This work was partially supported by the Argentinian grants PIP 2021 11220200100012
and PICT 2021-GRF-TII-00061 awarded by CONICET and ANPCYT, respectively. 
\end{acknowledgement}


\bibliographystyle{baaa}
\small
\bibliography{ref}

\end{document}